\documentclass{llncs}
\usepackage{llncsdoc}
\usepackage{graphicx}
\usepackage{subfigure}
\usepackage{makeidx}
\usepackage{hyperref}
\usepackage{algorithm}
\usepackage{algorithmic}
\usepackage{multirow}
\usepackage{amsmath}
\usepackage{amssymb}
\usepackage{xcolor}
\usepackage{epsfig}
\usepackage{graphics}

\begin{document}

%{\itshape Sample Output\/} (follows on the next page together with
%examples of the above run-in headings)
\newcounter{save}\setcounter{save}{\value{section}}
{\def\addtocontents#1#2{}%
\def\addcontentsline#1#2#3{}%
\def\markboth#1#2{}%
\title{Hybrid Parallel Bidirectional Sieve based on SMP Cluster}%\footnote{Term Project Final Report\\Student ID: 100520124 \\Presented \today}}

\author{Gang Liao \and Lian Luo \and Lei Liu}

\institute{Computer Science and Engineering Department,
 Sichuan University Jinjiang College, 620860
 Pengshan, China\\
 greenhat1016@gmail.com,mr.l172586418@gmail.com,cys19900611@gmail.com}

\maketitle

\begin{abstract}
In this article, hybrid parallel bidirectional sieve  is implemented by SMP Cluster, the individual computational units joined together by the communication network, are usually shared-memory systems with one or more multicore processor. To high-efficiency optimization, we propose average divide data into nodes, generating double-ended queues (deque) for sieve  that are able to exploit dual-cores simultaneously start sifting out primes from the head and tail.And each node create a FIFO queue as dynamic data buffer to ache temporary data from another nodes send to. The approach obtains huge speedup and efficiency on SMP Cluster.
\end{abstract}

\section{Introduction}
Research into questions involving primes continues today, partly driven by the importance of primes in modern cryptography. As our computational power increases, researcher often pays more attention to Data analysis, Climate modeling, Protein folding, Drug discovery etc. We can also exploit multicores to efficiency solve some problem in the field of number theory.

M.Aigner and G.M.Ziegler \cite{1} presented six quite different proofs of the infinitude of primes. Mills\cite{2} has shown that there is a constant $\Theta$ such that the function
$f(n)=[{\Theta^{3}}^{n}]$ generates only primes. The sieve of Eratosthenes-Legendre \cite{3} \cite{4} is an ancient algorithm for finding all prime numbers up to any given limit. In number theory, tests distinguishing between primes and composite integers will be crucial. The most basic primality test is trial division, which tells us that integer $n$ is prime if and only if it is not divisible by any prime not exceeding $\sqrt{n}$.

The computational complexity of algorithms for determining whether an integer is prime is measured in terms of the number of binary digits in the integer. The algorithm using trial divisions to determine whether an integer $n$ is prime is exponential in terms of the number of binary digits of $n$, or in terms of $\log_{2}n$ ,because $\sqrt{n} = {2}^{log_{2}{n/2}}$.

As n gets large, an algorithm with exponential complexity quickly becomes impractical. Leonard Adleman, Carl Pomerance, and Robert Rumely \cite{5} \cite{6} developed an algorithm that can prove an integer is prime using $(\log n)^{c log log log n}$ nit operations, where c is a constant. In 2002, M. Agrawal, N. Kayal, and N. Saxena \cite{7}, announced that they had found an algorithm ※PRIMES is in P§ that can produce a certificate of primality for an integer n using $O((log n) ^{12})$ bit operations.

Karl Friedrich Gauss conjectured that $\pi(x)$ increases at the same rate as the functions $\frac{x}{log x}$ and $Li(x) = \int_{2}^{x} \frac{dt}{log t}$. And the Prime Number Theorem that the ratio of $\pi(x)$ to $\frac{x}{log x}$ approaches 1 as $x$ grows without bound. One way \cite{11} to evaluate $\pi(x)$  only $O({x}^{\frac{3}{5}+e})$ bit operations without finding all the primes less than $x$ is to use a counting argument based on the sieve of Eratosthenes.

In this paper, Hybrid parallel bidirectional sieve  based on SMP Cluster is proposed to improve efficient and speedup. The result is proved to be effective by MPI and OpenMP \cite{8} \cite{9} \cite{10}. With Hybrid parallel, it has far-reaching significance in cryptography.

\section{Communication and Optimization}
ILP and TLP provide parallelism at a very low level, they are typically controlled
by the processor and the operating system, and isn't directly controlled by the programmer. Parallel hardware is often classified using Flynn's taxonomy, which distinguished between the number of instruction streams and the number of data streams a system can handle. A von Neumann system is classified as SISD. Vector processors and graphics processing units (GPU) are often classified as SIMD. MIMD execute multiple independent instruction streams, each of which can have its own data stream. Shared-memory or distributed-memory is typically MIMD. And most of the lager MIMD systems are hybrid systems (Fig.\ref{figc1}) in which a number of relatively small share-memory are connected by an interconnection network. In such systems, the individual shared-memory systems are sometimes called nodes.

\begin{figure}
\centering
\includegraphics[width=0.9\textwidth]{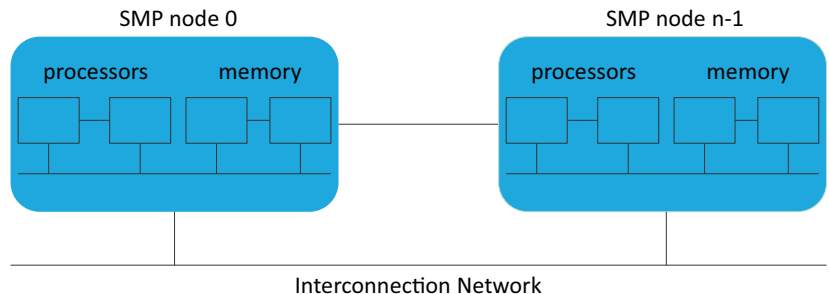}
\caption{SMP Cluster Architecture.}
\label{figc1}
\end{figure}

\subsection{Interconnection networks}
Currently the two most widely used interconnects on shared-memory systems are buses and crossbars \cite{15}. The key characteristic of a bus is that the communication wires are shared by the devices that are connected to it. Buses have the virtue of low cost and flexibility. Crossbars (Fig.\ref{figc2}) allow simultaneous communication among different devices, so they are much faster than uses. But the cost of the switches and links is relatively high.

Distributed-memory interconnects are often divided into two groups: direct interconnects and indirect interconnects. One measure of "number of simultaneous communications" or "connectivity" is bisection width. To understand this measure, imagine that the parallel system is divided into two halves, and each half contains half of the processors or nodes. An alternate way of computing the bisection width is to remove the minimum number of links needed to split the set of nodes into two equal halves.

\begin{figure}
\centering
\includegraphics[width=0.6\textwidth]{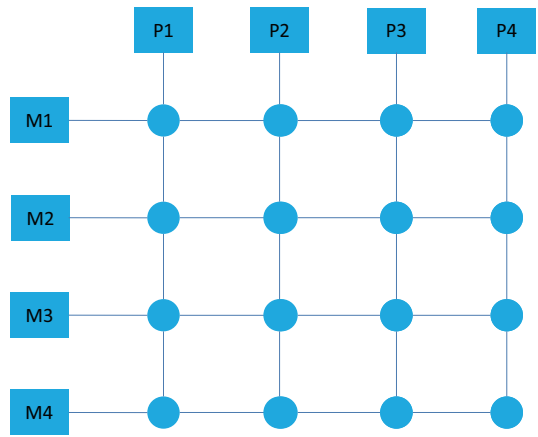}
\caption{Shared-memory system simultaneous memory access}
\label{figc2}
\end{figure}

The hypercube (Fig.\ref{figc3}) is a highly connected direct interconnect that has been used in actual system. A hypercube of dimension d has $p = {2}^d$ nodes, and a switch in a d-dimensional hypercube is directly connected to a processor and d switches. The bisection width of a hypercube is $\frac{p}{2}$.The switches support $1 + d = 1 + \log_{2}p$ wires. The hypercube is more powerful and expensive to construct.

\begin{figure}
\centering
\includegraphics[width=0.9\textwidth]{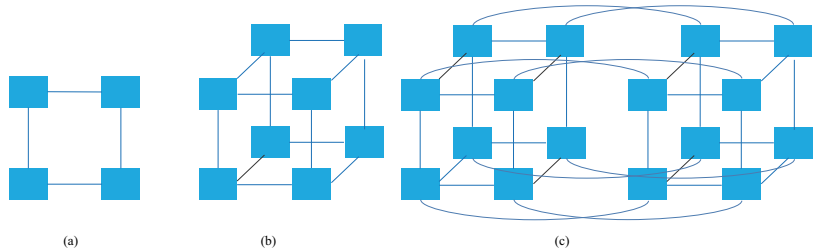}
\caption{(a) two-dimensional hypercube (b) three-dimensional hypercube (c) four-dimensional hypercube}
\label{figc3}
\end{figure}

The crossbar and the omega network are relatively simple examples of indirect networks. The omega network (Fig.\ref{figc4}) is less expensive than crossbar. The omega network uses $\frac{1}{2}plog_{2}(p)$ of the 2 x 2 crossbar switches, so it uses a total of ${2}plog_{2}(p)$ switches, while the crossbar users $p^{2}$.

\subsection{Hybrid Parallelism}
We define the speedup of a parallel program to be $S = \frac{T_{serial}}{T_{parallel}}$ . Then linear speedup has $S = P cores$, this value, $\frac{S}{P}$, is sometimes called the efficiency of the parallel program as follows:

\begin{equation}
E = \frac{S}{P} = \frac{\frac{T_{serial}}{T_{parallel}}}{P}
\end{equation}

Back in the 1960s, Gene Amdahl [13] that's become as Amdahl's Law:
 \begin{equation}
 S_{overall} = \frac{1}{(1-f)+\frac{f}{s}}
 \end{equation}

It means that unless virtually all of a serial program is parallelized, the possible speedup is going to be very limited-regardless of the number of cores available. A more mathematical version of this statement is known as Gustafson's Law \cite{14}.

Unfortunately, there are several mismatch problem between the (hybrid) programming schemes and the hybrid hardware architecture. Often, one can see in publications, that applications may or may not benefit from hybrid programming depending on some application parameters, e.g., in \cite{16}\cite{17}\cite{18} \cite{19}.

Polf Rabenseifner analyses strategies to overcome typical drawbacks of this easily usable programming scheme on systems with weaker inter-connects \cite{20}. Best performance can be achieved with overlapping communication and computation, but this scheme is lacking in ease of use. Often, hybrid MPI $+$ OpenMP programming denotes a programming style with OpenMP shared memory parallelization inside the MPI processes (i.e., each MPI process itself has several OpenMP threads) and communication with MPI between the MPI processes, but only outside of parallel regions.

This hybrid programming scheme will be named materonly in the following classification, which is based on the question, when and by which thread(s) the messages are sent between the MPI processes:

\begin{itemize}
\item[.] Pure MPI
\item[.] Hybrid MPI $+$ OpenMP
\item[.] Overlapping communication and computation
\item[.] Pure OpenMP
\end{itemize}

Overlapping of communication and computation is a chance for an optimal usage of the application itself, in the OpenMP parallelization and in the load balancing. It requires a coarse-grained and thread-rank-based OpenMP parallelization, the separation of halo-based computation from the computation that can be overlapped with communication, and the threads with different tasks must be load balanced.
Advantages of the overlapping scheme are:
\begin{itemize}
\item[.] the problem that one CPU may not achieve the inter-node bandwidth is no longer relevant as long as there is enough computational work that can be overlapped with the communication
\item[.] the saturation problem is solved as long as not more CPUs communicate in parallel than necessary to achieve the inter-node bandwidth
\item[.] the sleeping threads problem is solved as long as all computation and communication is load balanced among the threads.
\end{itemize}

\begin{figure}
\centering
\includegraphics[width=0.9\textwidth]{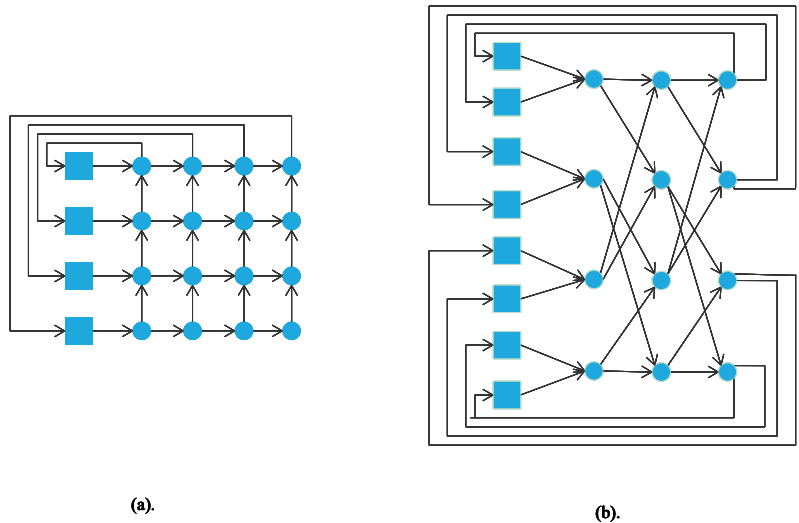}
\caption{ (a) Crossbar (b) omega network}
\label{figc4}
\end{figure}

\section{Bidirectional Sieve  Model}
Foster's methodology \cite{12} provides an outline of steps include
\begin{itemize}
\item[.] Partitioning.
\item[.] Communication.
\item[.] Agglomeration or aggregation
\item[.] Mapping for parallel programming
\end{itemize}

\subsection{Algorithm Design}
The sieve of Eratosthenes does so by iteratively marking as composite the multiples of each prime, starting with the multiples of 2 \cite{4}. We can exploit and improve the sieve of Eratosthenes based on SMP Cluster (Fig. \ref{figc5}). Assume that there are some disorder integers which the scale of $n$, and when each node sieve the integers in the block that the scale of $k$, it could achieve high-efficiency optimization. We conjectured that the SMP Cluster requires at least N nodes.The formula as follows:

\begin{equation}
N = \frac{n}{k}+(n \bmod k)\And{1}
\end{equation}

And each node generate one deque and do with dual-cores. One core is located in the head of the deque. On the contrary, the other one is located in the tail of the deque. It's easy to deduction the formula about the amount of cores($C_{cores}$) and deques($D_{deques}$):
\begin{equation}
C_{cores} = D_{deques} = {2}N
\end{equation}

There is another point that's worth considering. In most cases, the scale of node $N$ is not exactly equal $k$. We can deal with the state as follows Alg.\ref{alg1}:
\begin{algorithm}
\caption{the scale of node $N^{th}$} \label{alg1}
\begin{algorithmic}
\REQUIRE $K$ denote that the currency scale of node $N^{th}$
\ENSURE $k$ denote that the general scale of node
\IF{${0}\leq K \leq \frac{k}{2}$} \STATE Node N assign single core to right or left sieve
\ELSE \STATE Node N assign dual-cores to simultaneous bidirectional sieve
\ENDIF
\end{algorithmic}
\end{algorithm}

\begin{figure}
\centering
\includegraphics[width=0.8\textwidth]{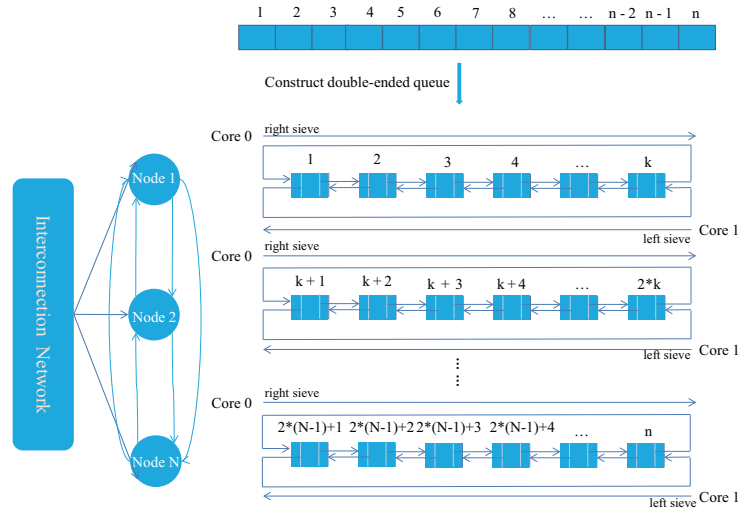}
\caption{Construct Bidirectional Sieve }
\label{figc5}
\end{figure}

And its flow diagram is shown in Fig.\ref{figc6}.
\begin{figure}
\centering
\includegraphics[width=0.6\textwidth]{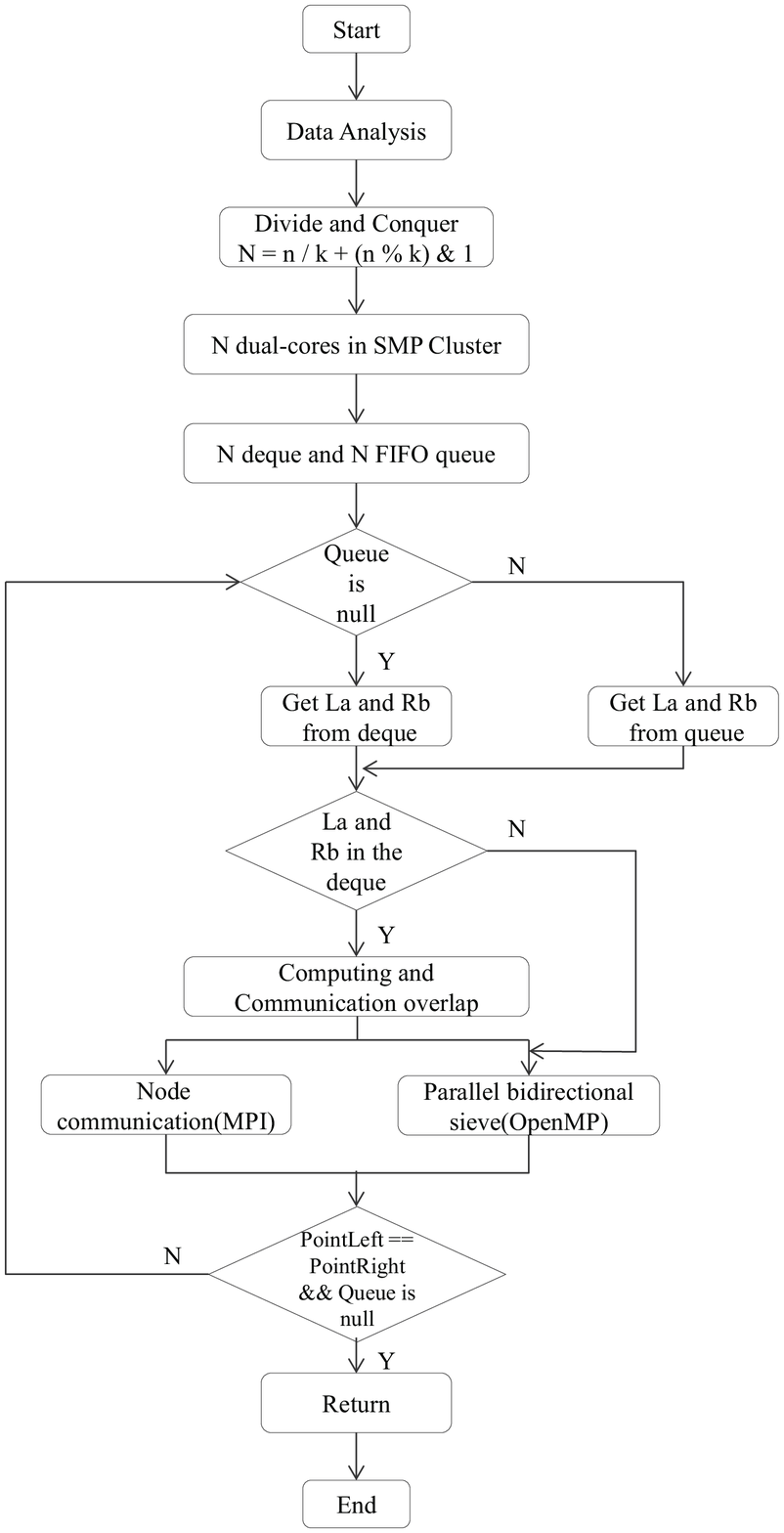}
\caption{High-level flow diagram of hybrid parallel bidirectional Sieve }
\label{figc6}
\end{figure}

\subsection{Primality Testing : Non-deterministic}
Primality testing of a number is perhaps the most common problem concerning number theory.The problem of detecting whether a given number is a prime number has been studied extensively but nonetheless,it turns out that all the deterministic algorithms for this problem are too slow to be used in real life situations and the better ones amongst them are tedious to code.But,there are some probabilistic methods which are very fast and very easy to code.Moreover,the probability of getting a wrong result with these algorithms is so slow that it can be neglected in normal situations.

All the algorithms which we are going to discuss will require you to efficiently compute $(a^b)\bmod c
$ (where a,b,c are non-negative integers).
A straightforward algorithm to do the task can be to iteratively multiply the result with $a$ and take
the remainder with $c$ at each step,this algorithm takes $O(b)$ time and is not very useful in practice.
We can do it $O(\log{b})$ by using what is called as exponentiation by squaring as follows:\\

$f(n) = $
$\begin{cases}
(a^{2})^{\frac{b}{2}}, & \mbox{if } b\mbox{ is even and b $>$ 0}\\
a(a^{2})^{\frac{b-1}{2}}, & \mbox{if } b\mbox{ is odd}\\
$1$,& \mbox{if } b\mbox{ $= 0$}
\end{cases}\\$

\begin{algorithm}
\caption{modulo(a,b,c) : Exponentiating by squaring to $(a^b)\bmod c$ } \label{alg2}
\begin{algorithmic}
\REQUIRE $x = 1,y = a$
\ENSURE $(a^b)\bmod c$

\WHILE{$b > 0$}
    \IF{$b \And 1$}
    \STATE $x = (x*y)\bmod c$
    \ENDIF

\STATE $y = (y*y) \bmod c$
\STATE $b>>=1$
\ENDWHILE
\RETURN $x \bmod c$
\end{algorithmic}
\end{algorithm}

Pierre de Fermat first stated the Fermat's Little Theorem in a letter dated October 18, 1640, to his friend and confidant Fr$\acute{e}$nicle de Bessy as the following \cite{7}:

\begin{equation}
a^p = a \pmod {p}
\end{equation}
or alternatively:
\begin{equation}
a^{p-1} = 1 \pmod {p}
\end{equation}

According to Fermat's Little Theorem\cite{7}, if $p$ is a prime number and a is positive integer less than $p$ ($a < p$),and then calculate $a^{p-1}\bmod p$. If the result is not 1, then by Fermat's Little Theorem p cannot be prime.The more iterations we do, the higher is the probability that our result is correct.

\begin{algorithm}
\caption{Fermat(p,iterations) : Fermat$'$s primality test} \label{alg3}
\begin{algorithmic}
\IF{$p = 1$}
    \RETURN $false$
    \ENDIF
\FOR{$i:=1$ to $iterations$}
\STATE $a = rand()\bmod (p-1) + 1$
    \IF{modulo(a,p-1,p)!=1 (Alg.\ref{alg2})}
        \RETURN $false$
    \ENDIF
\ENDFOR
\RETURN $true$
\end{algorithmic}
\end{algorithm}
Though Fermat is highly accurate in practice there are certain composite numbers $p$ known as Carmichael
numbers for which all values of $a<p$ for which $gcd(a,p)=1$,$(a^{p-1}) \bmod p=1$.And in that case,the
Fermat's test will return wrong result with very high probability.Out of the Carmichael numbers less than ${10}^{16}$,about $95\%$ of them are divisible by primes $<1000$.However,there are other improved primality tests which don't have this flaw as Fermat's(e.g.Rabin-Miller test\cite{21}\cite{22},Solovay-Strassen test \cite{23}).

\section{Performance Analysis}
\begin{figure}
\centering
\includegraphics[width=0.8\textwidth]{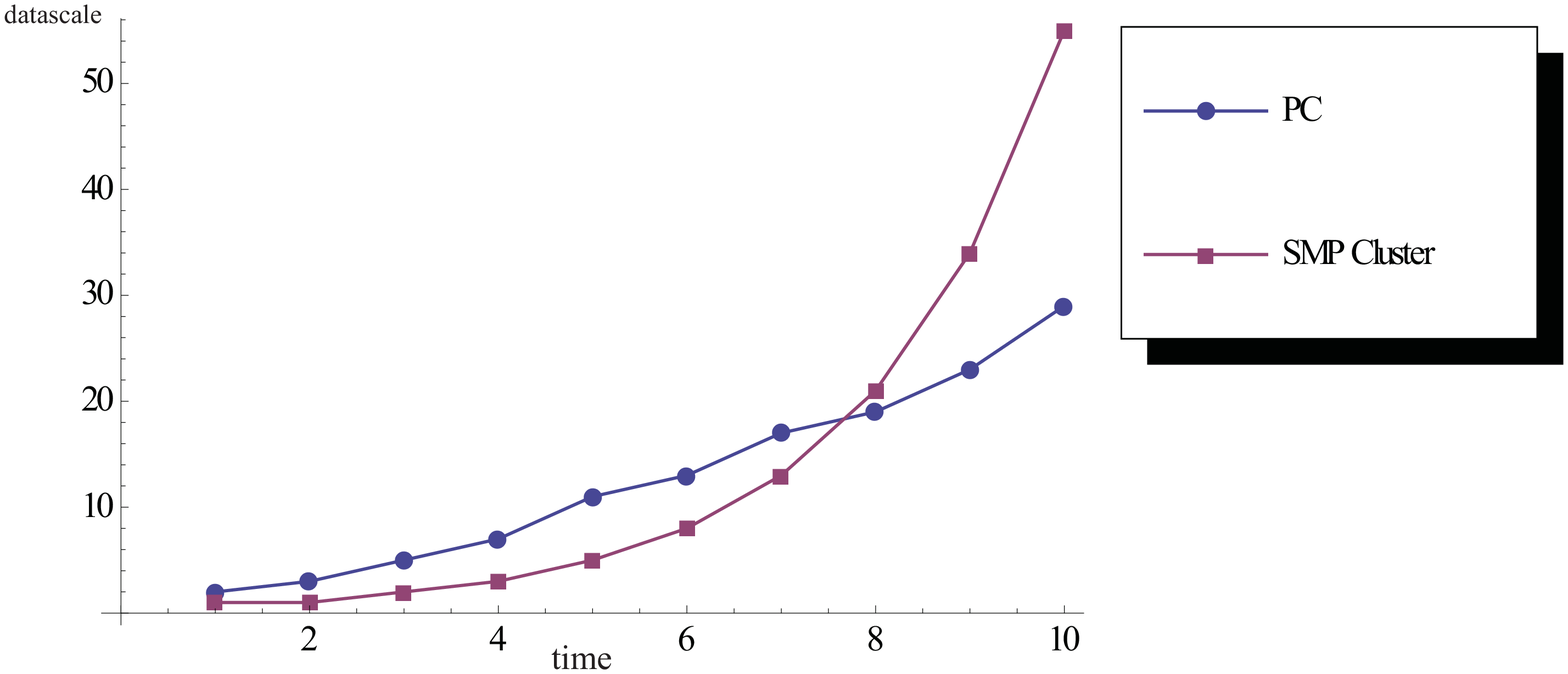}
\caption{statistics and analysis hybrid parallel bidirectional sieve with general method  }
\label{figc7}
\end{figure}
Different programming schemes on clusters of SMPs show different performance benefits or penalties in this paper.
Fig.\ref{figc7} summarizes the result of hybrid parallel bidirectional sieve .It's obvious that nodes communication would waste most of time when data scale is tiny.Even its slower than general method.However,
if  there are hyper-data scale,hybrid parallel show huge efficiency and optimization.Indeed,sometimes the waste of communication could be neglected.In that case,multicores parallelism  is effective approach to solve some problem in number theory.

To achieve an optimal usage of the hardware,one can also try to use the idling CPU's for other applications,especially low-priority single-threaded or multi-threaded non-MPI application if the
parallel high-priority hybrid application does not use the total memory of the SMP nodes.
\section{Conclusion}
In this study we haven shown that hybrid parallel on SMP cluster is an applicable method to implement bidirectional sieve . The analysis demonstrated that even hybrid parallel bidirectional sieve is
efficiency and optimization solution.

As our computational power increases,Most HPC system are clusters of shared memory nodes.Parallel programming must combine the distributed memory parallelization on the node inter-connect with shared memory parallelization inside of each node.And Each parallel programming schema on hybrid architecture has one or more significant drawbacks(e.g. sleeping-thread and saturation problem).
However,Hybrid parallel also has far-reaching significance in many fields(e.g.Cryptography,Data analysis, Climate modeling, Protein folding, Drug discovery).

We believe that hybrid parallel bidirectional sieve  can be properly modeled using techniques form number theory and this article is just an early trial of using hybrid parallelism to improve speedup and efficiency.


\begin{thebibliography}{[MT1]}
%
\bibitem[1]{1}
M. Aigner , G. M. Ziegler: Proofs from THE BOOK,3rd ed.,Springer-Verlag,Berlin,(2003)
%

\bibitem[2]{2}
W.H. Mills: A prime-representing function, Bulletin of the American Mathematical Society, Volume 53,604,(1947)
%

%
\bibitem[3]{3}
H.Halberstam , H.-E.Richert: Sieve Methods, Academic Press, London,(1974)
%

%
\bibitem[4]{4}
Horsley, Rev. Samuel, F. R. S.: The Sieve of Eratosthenes. Being an Account of His Method of Finding All the Prime Numbers, Philosophical Transactions (1683每1775), Vol. 62., pp. 327每347,(1772)
%

%
\bibitem[5]{5}
L.M. Adleman, C. Pomerance, R.S. Rumly, On distinguishing prime numbers from composite numbers, Annals of Mathematics, Volume 117 (1983).
%

%
\bibitem[6]{6}
 R. Rumely,:Recent advances in primality testing, Notices of American Mathmatical Society, Volume 30 , 475-477,(1983)
%
\bibitem[7]{7}
M.A Agrawal, N. Kayal, N. Saxena : PRIMES is in P, Department of Computer Science \& Engineering, Indian Institute of Technology, Kanpur, India,(2002)

%
\bibitem[8]{8}
R. Chandra, et al.: Parallel Programming in OpenMP, Morgan Kaufmann, San Francisco,(2001)
%
\bibitem[9]{9}
P. Pacheco: Parallel Programming with MPI, Morgan Kaufmann, San Francisco,(1997)
%
\bibitem[10]{10}
M.Quinn: Parallel Programming in C with MPI and OpenMP, McGraw-Hill Higher Education, Boston,(2004).
%
\bibitem[11]{11}
J.C. Lagarias , A.M. Odlyzko: New algorithm for computing PI(x), Bell Laboratories Technical Memorandum TM-82-11218-57.
%
\bibitem[12]{12}
I. Foster: Designing and Building Parallel Programs, Addison-Wesley, Reading, MA, 1995. Also available from http://www.mcs.anl.gov/~itf/dbpp/ (accessed 21.09.10)
%
\bibitem[13]{13}
G.M. Amdahl: Validity of the single processor approach to achieving large scale computing capabilities, in: Proceedings of the American Federation of Information Processing Societies Conference, vol. 30, issue 2, Atlantic City, NJ, 1967, pp. 483-485
%
\bibitem[14]{14}
J.L. Gustafson: Reevaluating Amdahl's law, Commun. ACM 31 (5) (1988) 532-533
%
\bibitem[15]{15}
Peter S. Pacheco: An introduction to PARALLEL PROGRAMMING, Elsevier (Singapore) Pte Ltd,(2011)
%
\bibitem[16]{16}
Georg Hager, Frank Deserno, Gerhaed Wellein: Pseudo-Vectorization and RISC Optimization Techniques for the Hitachi SR8000 Architecture, in High Performance Computing in Science and Engineering in Munich '02, Springer-Verlag Berlin Heidelberg, (2003)
%
\bibitem[17]{17}
D. S. Henty: Performance of hybrid message-passingand shared-memory parallelism for discrete element modeling, in Proc. Supercomputing'00, Dallas, TX, (2000).
%
\bibitem[18]{18}
Richard D. Loft, Stephen J. Thomas, John M. Dennis: Terascale spectral element dynamical core for atmospheric general circulation models, in proceedings, SC 2001, NOW. 2001, Nov. 2001, Denver, USA. www.sc2001.org/papers/pap.pap189.pdf
%
\bibitem[19]{19}
Gerhard Wellein, GeorgHager, Achim Basermann,  Holger Fehske: Fast sparse matrix-vector multiplication for TeraFlop/s computers, in proceedings of VECPAR'2002, 5th Int'l Conference on High Performance Computing and Computational Science, Porto, Portugal, June 26-28, 2002, part I, pp 57-70. http://vecpar.fe.up.pt/
%
\bibitem[20]{20}
Rolf Rabenseifner: Hybrid Parallel Programming: Performance Problems and Chances, in proceeding of the 45th CUG Conference 2003, Columbus, Ohio, USA, May 12-16,2003, www.cug.org
%
\bibitem[21]{21}
Miller, Gary L.:Riemann's Hypothesis and Tests for Primality, Journal of Computer and System Sciences 13 (3): 300每317, doi:10.1145/800116.803773,(1976)
%
\bibitem[22]{22}
Rabin, Michael O.:Probabilistic algorithm for testing primality, Journal of Number Theory 12 (1): 128每138, doi:10.1016/0022-314X(80)90084-0,(1980)
%
\bibitem[23]{23}
Solovay, Robert M.Strassen, Volker.:A fast Monte-Carlo test for primality". SIAM Journal on Computing 6 (1): 84每85. doi:10.1137/0206006,(1977)
\end{thebibliography}
\end{document}